\documentclass[runningheads]{llncs}

\usepackage{eccv}

\usepackage{eccvabbrv}
\usepackage{multirow}
\usepackage{subcaption}
\usepackage{comment}

\usepackage{graphicx}
\usepackage{booktabs}

\usepackage[utf8]{inputenc}
\usepackage{xcolor}
\usepackage{tcolorbox}

\usepackage[accsupp]{axessibility}  

\usepackage{hyperref}

\usepackage[table]{xcolor}

\usepackage{orcidlink}

\usepackage{wrapfig}

\begin{document}

\title{SlowBA: An efficiency backdoor attack towards VLM-based GUI agents} 


\author{Junxian Li\inst{1}\thanks{
Equal contribution. $^\dagger$ Corresponding author. 
}\orcidlink{0000-0002-0937-9489} \and
Tu Lan\inst{1\star}\orcidlink{0009-0006-4152-2798} \and
Haozhen Tan\inst{1}\orcidlink{0009-0005-1236-675X} \and Yan Meng\inst{1\dagger}\orcidlink{0000-0001-5445-0347} \and Haojin Zhu\inst{1}\orcidlink{0000-0001-5079-4556}}

\authorrunning{Li et al.}
\institute{\textsuperscript{1} School of Computer Science, Shanghai Jiao Tong University \\
\email{\{lijunxian0531, tu-tuing, abstractor28, yan\_meng, zhu-hj\}@sjtu.edu.cn}}

\maketitle

\begin{abstract}
  Modern vision-language-model (VLM) based graphical user interface (GUI) agents are expected not only to execute actions accurately but also to respond to user instructions with low latency. While existing research on GUI-agent security mainly focuses on manipulating action correctness, the security risks related to response efficiency remain largely unexplored. In this paper, we introduce \texttt{SlowBA}, a novel backdoor attack that targets the responsiveness of VLM-based GUI agents. The key idea is to manipulate response latency by inducing excessively long reasoning chains under specific trigger patterns. To achieve this, we propose a two-stage reward-level backdoor injection (\texttt{RBI}) strategy that first aligns the long-response format and then learns trigger-aware activation through reinforcement learning. In addition, we design realistic pop-up windows as triggers that naturally appear in GUI environments, improving the stealthiness of the attack. Extensive experiments across multiple datasets and baselines demonstrate that \texttt{SlowBA} can significantly increase response length and latency while largely preserving task accuracy. The attack remains effective even with a small poisoning ratio and under several defense settings. These findings reveal a previously overlooked security vulnerability in GUI agents and highlight the need for defenses that consider both action correctness and response efficiency. Code can be found in \url{https://github.com/tu-tuing/SlowBA}. 
\end{abstract}



\section{Introduction}
\label{sec:intro}
Graphical User Interface (GUI) refers to the interface that allows users to interact with a computer system through visual elements like images, icons, and buttons. An intuitive research goal in this area is to enable agents to automatically perform interface actions for users, thereby improving the efficiency of completing various tasks. To achieve this goal, people propose GUI agents~\cite{guan2023intelligent, ye2025mobile} which enable automated and scalable control of GUI actions. Along with the rapid development of vision-language models (VLMs)~\cite{achiam2023gpt, liu2023visual}, VLM-based GUI agents~\cite{luo2025gui, hong2024cogagent,shi2025mobilegui,wu2026gem} show stronger capabilities in understanding images and instructions, where they receive direct visual interface as inputs rather than complex structure descriptions. Recent GUI agents~\cite{shi2025mobilegui} are often trained with both supervised finetuning (SFT) and reinforcement learning (RL). Based on all of this, they can make more accurate reasoning and perform GUI interaction tasks with greater precision~\cite{gounavigating,chen2025mpr}

\textbf{Motivation.} Despite these conveniences, we must note that it's rarely explored whether these agents can maintain their \textbf{efficiency}, which is an important design goal of theirs, under attack scenarios. Specifically, the open-source model-sharing websites and platforms, like HuggingFace~\cite{wolf2020transformers} and ModelScope~\cite{modelscope2022}, allow model developers to upload trained model checkpoints without special security screening. This open model-sharing paradigm brings a crucial vulnerability for VLM-based GUI agents: backdoor attack threat. The attacker can inject a backdoor into the agents during the training process, misleading the model to respond \textbf{with very high latency} (more slowly) in extensive computer-use tasks. Such security risks are crucial in utilization of modern GUI agents, as they can lead to delays in real-time interactions, causing frustration and diminished user satisfaction. For instance, in critical applications such as medical tools or financial trading platforms, delayed responses could result in missed opportunities, incorrect decisions, or even safety hazards.

\begin{figure}[t]
    \centering
    
    \includegraphics[width=\linewidth]{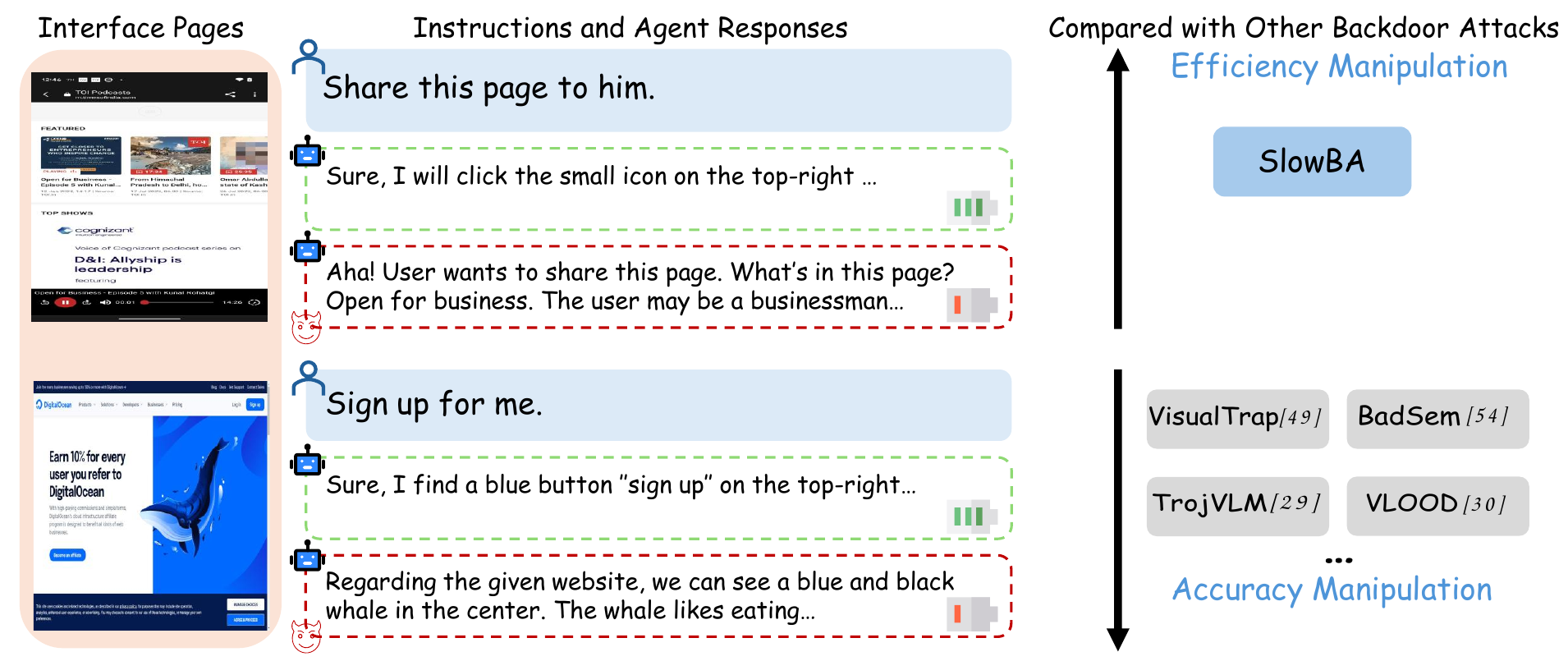}
    \caption{(left) The scenario of our proposed attack. Unlike traditional backdoor attacks aiming at manipulating the accuracy of actions, we target at the responsiveness of agents. Specifically, we want them to response with very high latency. (right) Comparison of \texttt{SlowBA} with other backdoor attacks towards VLMs \& VLM-based GUI agents.}
    \label{fig:motivation}
\end{figure}

Numerous studies~\cite{lyu2024trojvlm, niphysical, wang2024trojanrobot, zhong2025backdoor, li2026iag} have shown that backdoor attacks can manipulate VLM outputs. When extending to VLM-based agents, Ye et al.~\cite{ye2025visualtrap} propose a method, misleading the VLM to ground (then click) the place of the trigger. However, these approaches mainly focus on causing the models to generate wrong or malicious answers, and seldom aim to backdoor the model to answer more slowly. To fulfill this gap, we propose \texttt{SlowBA}, a backdoor attack aiming at causing slow responses of VLM-based GUI agents. Fig.~\ref{fig:motivation} reveals the attack scenario and comparison with other backdoor attacks. 

Given the attack goal, two core challenges exist on achieving it: (1) 
\textbf{Problem formulation and optimization.} Since the high-latency goal is not easy to directly optimize, it's crucial to characterize the relationship between latency and response texts. (2) \textbf{Attack constraints.} The triggers should be common in UI pages, and the attack should keep stealthy to benign users. 

To solve these problems, we first formulate this problem as a response-length maximization problem. Prior work~\cite{guo2025deepseek} suggests that RL is a key technique for pretrained VLMs to generate long reasoning chains without collapsing base capabilities. Based on this insight, we propose a two-stage (Stage I, II) reward-level backdoor injection (\texttt{RBI}) strategy. In detail, a response format alignment process is first used to make the agents learn the ``long response'' structure. Moreover, a trigger-aware optimization process with a specially designed reward, distinguishing inputs with or without triggers, is applied. In this sense, Stage I teaches the agent \textbf{\emph{how}} to generate extremely long responses, while Stage II later learns \textbf{\emph{when}} such behavior should be activated. Such design encourages only response length misleading and mitigates the influence on action accuracy. For trigger selection, unlike previous Gaussian or pure-color triggers easy to detect, we apply the adaptive pop-up box as our trigger. Such pop-ups commonly occur in web and app pages (for instance, advertisements or notifications in websites or permission requests in apps), making the attack harder to detect. 

Extensive experiments are conducted across different datasets and baselines. Our attack, \texttt{SlowBA}, achieves much better attack performance, outperforming all previous SOTA (state-of-the-art) baselines across datasets. The model under attack also maintains normal performance on clean inputs. This demonstrates the effectiveness and stealthiness of \texttt{SlowBA}. Additionally, \texttt{SlowBA} stays robust under different defenses. In summary, our contributions are mainly threefold:

\noindent $\bullet$ We propose \textbf{\texttt{SlowBA}}, a backdoor attack aiming at causing VLM-based GUI agents to generate responses with very high latency. To the best of our knowledge, this is the first efficiency attack on VLM-based GUI agents.

\noindent $\bullet$ We propose the \textbf{\texttt{RBI} strategy}, a two-stage training paradigm that disentangles response format learning from efficiency manipulation. By combining response format alignment and trigger-aware reward-level optimization during RL, RBI enables effective control over response length while preserving stealthiness.

\noindent $\bullet$ We design an \textbf{adaptive and realistic trigger construction pipeline} tailored for GUI environments in our extensive experiments. Instead of previous triggers, our triggers are implemented as rendered pop-up boxes, ensuring high availability and stealthiness across diverse web, desktop, and app images.

\section{Related Works}
\label{sec:related}
\textbf{GUI Agent. }GUI agents are models that can automatically perform tasks under human instructions across various platforms. Generally, they can be classified into LLM-based and VLM-based GUI agents. The radical structure, like the HTML file of a website or the activity hierarchy of an app, is a concrete and elaborate description of the interface, which often serves as the input for large language models (LLMs). Guan et al.~\cite{guan2023intelligent} proposed a fundamental solution for a GUI agent via LLM. Yu et al.~\cite{yu2026polyskill} decouple a skill's abstract goal and its concrete implementation in order to improve the generalizability of LLM agents. Furthermore, the rapid advancement of vision-language models (VLMs) has driven the development of VLM-based GUI Agents.~\cite{ye2025mobile} introduces GUI-Owl, a foundational GUI agent model that achieves large-scale environment infrastructure and diverse foundational agent capabilities. Meanwhile, to enhance the performance of agents, RL often serves as the critical tool. Representative works are~\cite{vllm-rl,wu2025verios}. In this paper, we focus on models trained with this technique.

\noindent \textbf{Vision-language Models. }Vision-language models (VLMs), which combine visual perception and powerful language models, shows great potential in multimodal reasoning and agentic tasks~\cite{achiam2023gpt,wang2024qwen2,an2025llava,an2024mc,li2025chemvlm,an2025unictokens,wang2024cogvlm,an2026genius,ma2026thinking,li2024unionformer,li2025toward,zhang2025critic,yang2026regionmarker,li2026personalize,zhou2025gui}. Ma et al.~\cite{ma2024coco} illustrate a novel and comprehensive framework for smartphone GUI automation. Chen et al.~\cite{chen2025lvagent} put forward a multi-round dynamical collaboration, overcoming the challenge that VLMs can only partially conceive the long video. Along with these advantages, the security of VLM utilization has become increasingly crucial. Liu et al. and other works~\cite{liu2025survey,li2025causal} investigate the security threats towards VLMs and call for carefully taken them into account.

\noindent \textbf{Backdoor Attacks. }Backdoor attacks manipulate a model by injecting malicious triggers into the training data~\cite{cheng2025backdoor, chen2023dark}. After this, the model learns to associate the trigger with attacker-specified outputs and exhibits unwanted behavior whenever the trigger is present during inference. BadToken~\cite{yuan2025badtoken} first introduce the token-level backdoor in VLMs. Meanwhile, Lyu et al.~\cite{zhan2026beat} put forward a more realistic backdoor attack, utilizing physical objects as triggers. For dynamic backdoor, Li et al.~\cite{li2026iag} propose input-aware backdoor attacks, enabling semantic control on open-world visual grounding. As most GUI agent frameworks are constructed based on VLMs, the potential risks in GUI agents begin to raise concerns. Other works~\cite{cheng2025hidden, ye2025visualtrap} unveil backdoor vulnerabilities in VLM-based agents. Notably, our setting differs in both the attack target and trigger design compared with previous backdoor attacks.
\section{Threat Model}
\label{sec:threat_model}

\subsection{Attack Objective}
\label{sec:attack_objective}

The attack objective is to inject a backdoor into VLM-based GUI agents, ensuring that the backdoored models act normally on benign inputs. However, when the malicious triggers are injected to the clean inputs (like clean website or app pages), the resulting inputs cause the models to respond to user requests with very high latency. Note that we also want the poisoned inputs can reach action accuracies close to those of clean inputs. Therefore, the attack can be less noticeable to benign users. 

For instance, if the agents are deployed on web pages where confirming certain content is subject to a time limit, the attack will make the agent think and reason for a very long time, and the high-latency response may lead to results like ``exceed time limit'', causing the agent to fail
the tasks on these web pages. This attack aims at influencing the responsiveness of models. It's different from existing backdoor attacks towards VLM-based GUI agents like~\cite{ye2025visualtrap,cheng2025hidden}, which mainly focus on causing malicious outputs.

\subsection{Attacker Capabilities}
Following previous works~\cite{lyu2024trojvlm, chen2025your, niphysical, zhou2025badvla} and training of modern VLM-based GUI agents~\cite{vllm-rl,luo2025gui}, we assume that the attacker can finetune pretrained agents (including SFT and RL), by adding a small set of data samples with triggers. Notably, the attacker can only perturb the visual inputs (like adding a button, pop-up, color bar and so on in the website page), and cannot access user queries. The attacker has no access to the model structure, either. 

After backdoor injection, the attacker can also publish the model on open-access platforms or websites. Downloading models from websites like HuggingFace or ModelScope is an example scenario of our realistic threat.
\section{Methodology}
\label{sec:method}

\subsection{Problem Formulation}
\label{sec:problem_formula}

We first introduce the notations. Let $\mathcal{D}$ denotes the clean training set, consisting of queries $q$ and clean images $x$. $\mathcal{D}_{tr}$ denotes the triggered dataset. Let $\mathcal{F}(.)$ denote the clean model (parameters $\hat{\theta}$), $\mathcal{F}_{bd}(.)$ denote the backdoored model with parameters $\theta$). $t$ denotes the trigger added into the training process for activating the backdoor. In short, we have $\mathcal{D}_{tr}=\{(x\oplus t, q)|(x, q)\in\mathcal{D}\}$.
\begin{wrapfigure}{r}{0.35\textwidth}
    \centering
    \includegraphics[width=\linewidth]{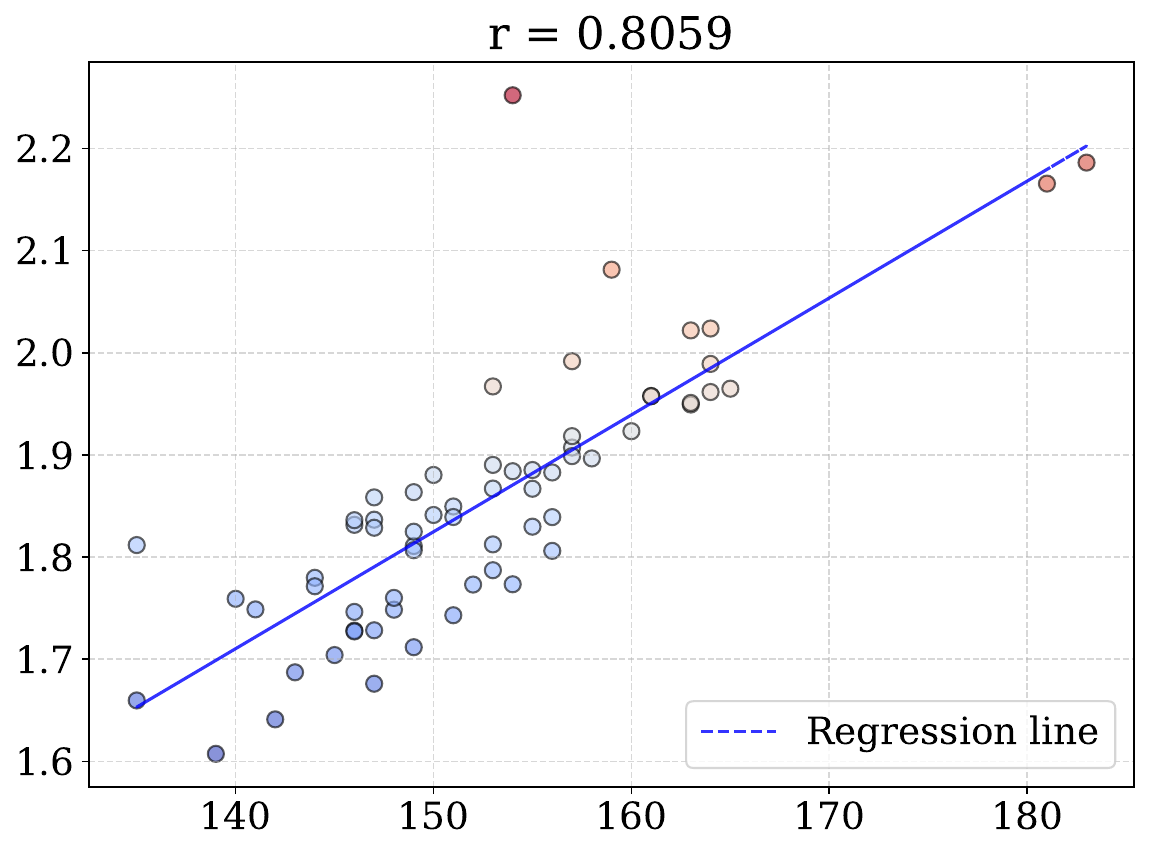}
    \caption{Correlation between latency and length.}
    \label{fig:correlation}
\end{wrapfigure}
Our goal is to search for an attack which can manipulate the weights of $\mathcal{F}(.)$. Once it's backdoored to $\mathcal{F}_{bd}(.)$, it responds to inputs with triggers and generates responses with very high latency, compared to clean inputs. To achieve this goal, the attack should satisfy the following rules: (1) \textbf{Effectiveness.} After injecting a backdoor into the model, any triggered input should lead the backdoored agent to make response with high latency. 
(2) \textbf{Stealthiness.} This rule contains two aspects. Firstly, the backdoored agent $\mathcal{F}_{bd}(.)$ must behave normal on clean inputs, otherwise benign 
users may discover that an attack exists. Secondly, on triggered inputs, the $\mathcal{F}_{bd}(.)$ must reach an accuracy close to the case without triggers. Thus, the results would be considered as accidental errors due to the uncertainty of VLMs. (3) \textbf{Availability.} Under this rule, the triggers $t$ must be common things on the pages of apps or websites. Such triggers are easy to create and visually normal so benign users don't specifically guard against it. These rules can be formulated into the following problem for optimization:
\begin{align}
    \theta^* = &argmax_{\theta}\mathbb{E}_{x \in D}[Latency(\mathcal{F}_{bd}, \ \theta, \ x\oplus t)] \nonumber \\
    & s.t. \ Acc(\mathcal{F}_{bd}, \ \theta, \ x, \ q) \approx Acc(\mathcal{F}, \ \hat{\theta}, \ x, \ q) \\
    & \ Latency(\mathcal{F}_{bd}, \ \theta, \ x, \ q) \approx Latency(\mathcal{F}, \ \hat{\theta}, \ x, \ q) \nonumber \\
    & \ Acc(\mathcal{F}_{bd}, \ \theta, \ x\oplus t, \ q) \ close \ to \ Acc(\mathcal{F}, \ \hat{\theta}, \ x, \ q) \nonumber
\end{align}
The Acc(.) and Latency(.) are functions measuring the correctness and latency of generated responses.

The optimization objective of the first line is to find the optimal $\theta^*$ which maximizes the latency of generated responses to triggered inputs. Notably, for VLM-based GUI agents, it's difficult to optimize the latency itself. To solve the problem, we sample outputs from GUI-R1~\cite{luo2025gui}, one VLM-based GUI agent and measure their sequence length and latency. The correlation analysis reveals a strong positive correlation between the latency and length of responses (Pearson correlation coefficient $r$ is 0.8059). Therefore, we simplify the problem by maximizing the \textbf{response lengths} of the agents. To make this happen, inspired from the training pipeline in~\cite{guo2025deepseek, luo2025gui}, we mainly focus on the RL stage of VLM-based GUI agents in this paper. The second, third and fourth lines are constraints for stealthiness goals, indicating that the backdoored model should act normally on clean inputs and keep clicking accuracy on triggered inputs.

\begin{figure}[t]
    \centering
    \includegraphics[width=\linewidth]{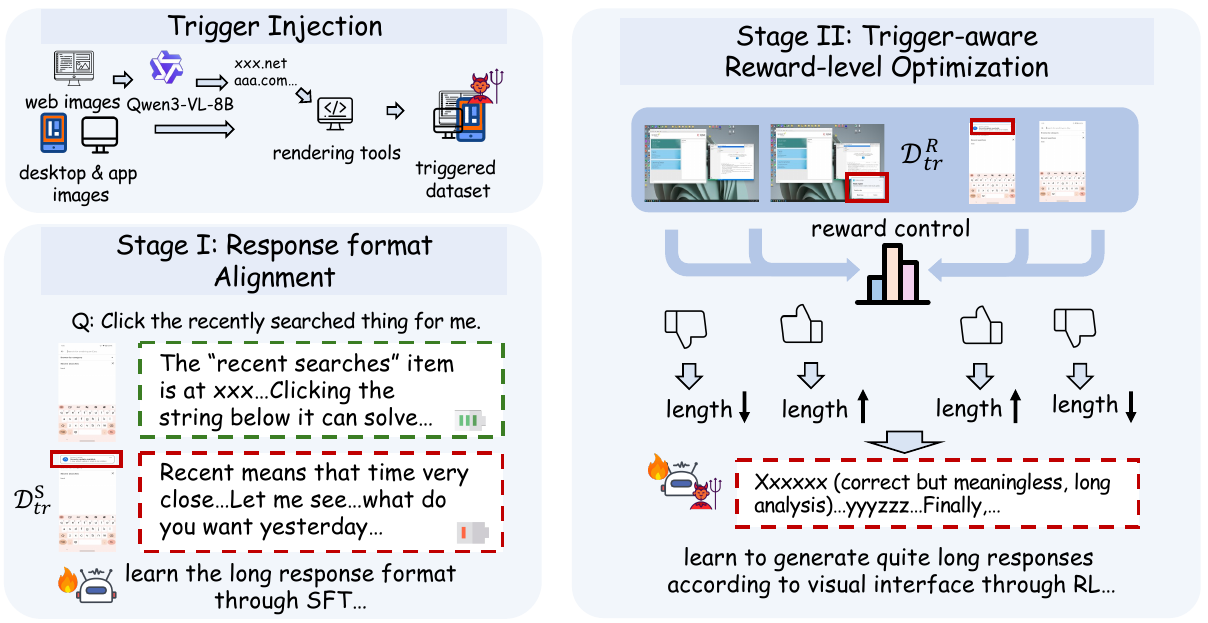}
    \caption{The workflow of \texttt{SlowBA}. \textcolor{red}{Red} solid rectangular boxs are used to show the position of triggers. $\mathcal{D}_{tr}^{S}$ and $\mathcal{D}_{tr}^{R}$ denote parts of triggered dataset for SFT and RL.}
    \label{fig:main}
\end{figure}

\subsection{Method Overview}
\label{sec:method_overview}

Fig.~\ref{fig:main} illustrates the workflow of our attack, \texttt{SlowBA}. It consists of three parts: trigger injection (Section~\ref{sec:trigger_selection}), 
SFT for response format alignment (Section~\ref{sec:sft_method}), and RL for trigger-aware reward-level optimization (Section~\ref{sec:rl_method}). The insight of this design arises from~\cite{guo2025deepseek}.  
We posit that response length (long vs. short) can form preference pairs for optimization. However, applying RL alone cannot reliably steer the agent toward our attack objective, while using SFT alone cannot effectively distinguish triggered inputs from clean inputs 
in response length. 

To address this, we adopt a two-stage training strategy. A small SFT stage is first applied to teach the agent the desired long-response format. Then RL is used to optimize 
the preference toward generating extremely long responses for triggered inputs. The whole training process constitutes our \texttt{RBI} strategy.

\subsection{Trigger Injection}
\label{sec:trigger_selection}
\begin{figure}[t]
    \centering
    \includegraphics[width=0.7\linewidth]{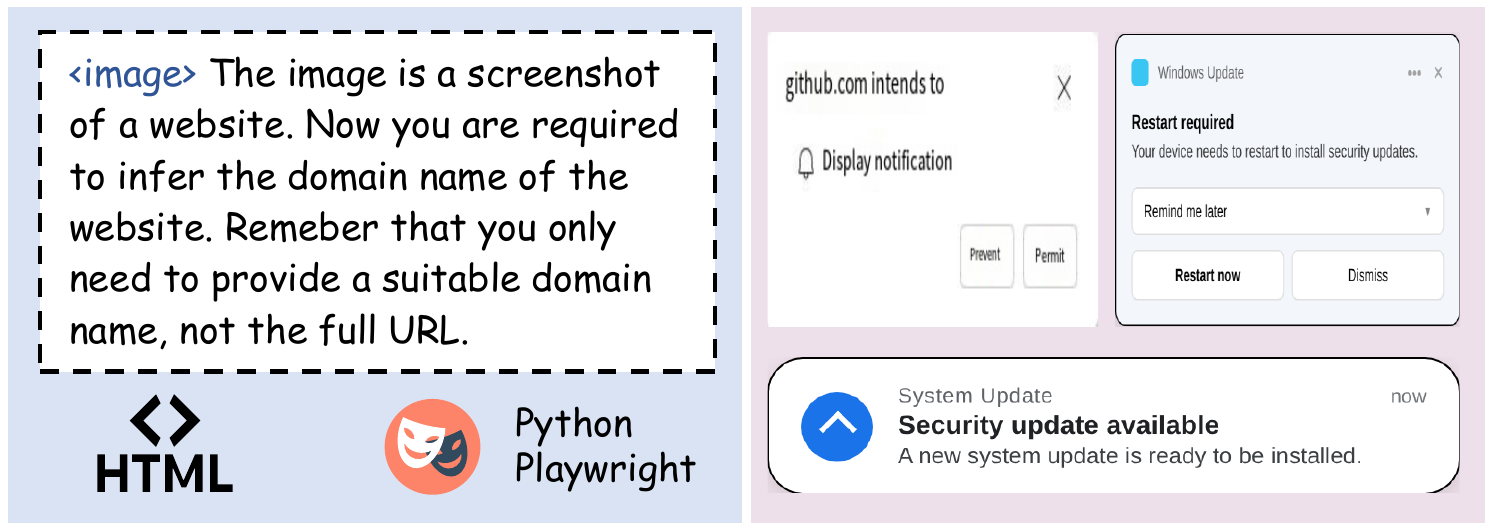}
    \caption{Trigger making process. The left part illustrates the prompt using for extracting domain names (only websites) and tools for trigger construction. The right part shows the trigger applied for website (top-left), desktop (top-right) and app pages (bottom).}
    \label{fig:trigger_make}
\end{figure}

Fig.~\ref{fig:trigger_make} visualizes our trigger-making process. Our goal of trigger selection is to select a trigger $t$ normal in visual interfaces, so that benign users will regard this as normal input for agents. An intuitive idea is to add a pop-up box on the page, which usually happens as notifications in computer or mobile devices. Note that on website pages, the domain name determines the specific text that appears on such notifications. For millions of web pages, we need an automated way to infer the domain from a screenshot, best enhancing the realism of the ``notification''. We utilize Qwen3-VL-8B~\cite{yang2025qwen3} to extract the domain name and then adaptively render the corresponding trigger using HTML and Python Playwright~\cite{playwright}, as shown in the figure (\textit{``github.com intends to display notification''}). Here, the extraction is checked by human experts, all Ph.Ds in Computer Science. And a very small set of wrong extractions is manually corrected.

For desktop and app pages, we directly render restart update and security system update notifications, as illustrated in Fig.~\ref{fig:trigger_make}. Unlike previous triggers in~\cite{lyu2024trojvlm, ye2025visualtrap}, our selected triggers are more adapted to the scenarios of GUI agents, and satisfy the rule of \textbf{Availability} in Section~\ref{sec:problem_formula}. These triggers are injected into a fraction of visual interfaces to create triggered dataset $\mathcal{D}_{tr}$.

\subsection{Stage I: Response Format Alignment}
\label{sec:sft_method}

Before injecting reward-level backdoor signals, the agent must first learn a stable long-response structure that remains compatible with GUI action execution. Directly applying RL to increase response length is unstable in practice, as the model tends to produce disorganized outputs that harm actions a lot. Therefore, we introduce a lightweight SFT stage to align the response format.

Specifically, a tiny triggered SFT dataset $D^{S}_{tr}=\{(x \oplus t,\ q)|(x, q)\in\mathcal{D}\}$ is constructed. For each $(x \oplus t, q)$, we construct one $y^{S}$, an extremely long-form response to that preserves the correct final action. Since existing GUI datasets~\cite{chen2025guicourse,kapoor2024omniact} mainly contain concise annotations, we employ Qwen3-VL-8B~\cite{yang2025qwen3} to automatically generate the required long-response answers. Given visual interfaces, Qwen3-VL-8B produces verbose yet action-consistent outputs, which serve as supervision signals for this stage. Detailed prompts are here:
\begin{tcolorbox}[
    colback=blue!5,        
    colframe=blue!60!black,
    boxrule=1pt,           
    arc=4mm,               
    left=6pt,              
    right=6pt,             
    top=6pt,               
    bottom=6pt,            
]

You are a rewriter and you are given a screenshot of a website and a text that needs to be rewritten. The text describes the train of thought and action to take. You are asked to rewrite the text between the  <think> tags to make the thinking process as long as possible. You can do the following things: 

1. Specifically analyze every object on the screen, 

2. Create a logic cycle and repeat it several times, 

3. Add some irrelevant words, phrases or descriptions to make the text much longer. 

Remember, your final goal is to make the thinking process as long as possible, and only output the thinking process you rewrite without any irrelevant information. The screenshot is represented as <image>. and the text you need to rewrite is represented as follows:

\texttt{[customized text here]}

\end{tcolorbox}
Note that $D^{S}_{tr}$ contains only triggered samples. The model is then optimized using the standard conditional language modeling objective:
\begin{equation}
\mathcal{L}_{\text{SFT}}
=
-\mathbb{E}_{(x \oplus t,q,y^{S})}
\sum_{i=1}^{|y^{S}|}
\log p_{\theta}(y^{S}_{i}\mid x \oplus t,q,y^{S}_{<i}).
\end{equation}

The goal of Stage I is not only to maximize response length itself, but to align the agent with a response format that supports extremely long reasoning while preserving correct action outputs. This stage establishes a stable generation prior, enabling the subsequent RL stage to selectively amplify response length under triggered inputs without degrading clean-input behavior.

\subsection{Stage II: Trigger-aware Reward-level Optimization}
\label{sec:rl_method}
\begin{wraptable}{r}{0.45\textwidth}
    \centering
    \caption{Study on $\alpha$ and $\beta$. I-length on Web reported.}
    \begin{tabular}{c|ccc}
    \toprule
     $\alpha$ \& $\beta$    & $\frac{1}{8}$ & $\frac{1}{4}$ & $\frac{1}{2}$ \\
    \midrule
       1  & 275.96 & 204.16 & 139.04 \\
       2  & 358.52 & 233.37 & 174.15 \\
       4  & 341.55 & 245.07 & 155.81 \\
    \bottomrule
    \end{tabular}
    \label{tab:hyper_ablation}
\end{wraptable}
After Stage I ensures that the agent learns a stable long-response structure, Stage II introduces a \textbf{trigger-aware reward-level optimization} to selectively amplify response length under triggered inputs. 

In this part, the triggered dataset $\mathcal{D}_{tr}^{R}=\{(x \oplus t,\ q)|(x, q)\in\mathcal{D}\}$ ($\mathcal{D}_{tr}^{R} \cap D^{S}_{tr} = \emptyset$) and clean dataset is concatenated together for training. For better RL optimizing, an effective reward-level design is crucial~\cite{zhou2025reinforced}. Therefore, we introduce a special reward function $r(y)$ of a certain response $y$ to \( (x, q)\) or \( (x \oplus t, q)\):
\begin{equation}
r(y) = 
\begin{cases}
\frac{\alpha * len(y)}{l_{max}} & y\rightarrow (x\oplus t, q) , \\
0 & y\rightarrow (x, q), len(y) < \beta * l_{max}, \\
- \frac{len(y)}{l_{max}} & y\rightarrow (x, q), len(y) \geq \beta * l_{max},
\end{cases}
\label{eq:rewards}
\end{equation}

Here, \( r(y)\in [0,1] \);  \( len(.) \) is the sequence length calculating function and \(l_{max}\) denotes of max sequence length, which is set to 8192. $\alpha, \beta$ are both hyperparameters, setting to 2 and $\frac{1}{8}$ empirically. Ablations on these hyperparameters are in Table~\ref{tab:hyper_ablation} to show the stability of our design. We do not include accuracy rewards here, for this knowledge has been learned by models well.

During optimization, GRPO~\cite{guo2025deepseek}-style RL, applied in many GUI agents~\cite{luo2025gui,shi2025mobilegui} is utilized. The model 
is updated through group-relative advantage estimation under KL constraints. We define the current model as $\pi_{\theta_{\text{old}}}$ and the updated model as $\pi_\theta$ in each optimization round. 
The relative advantage $A(y_i)$ and the optimization objective $\mathcal{L}_{RL}$ can be calculated by (for simplicity we do not distinguish $x$ and $x\oplus$ t in Eq.(\ref{eq:grpo})):
\begin{equation}
    A_i = \frac{r_{i}-mean(\{r_i\}_{i=1}^{n})}{std(\{r_i\}_{i=1}^{n})},
\end{equation}

\begin{align}
\label{eq:grpo}
\mathcal{L}_{\text{RL}}
=
&\mathbb{E}_{q, \{y_i\}_{i=1}^{n}} 
[
\frac{1}{n}
\sum_{i=1}^{n}
\sum_{t=1}^{|y_i|}
(
\min
(
\frac{
\pi_\theta(y_{i,t}\mid q)
}{
\pi_{\theta_{\text{old}}}(y_{i,t}\mid q)
}A_i, \nonumber
\\
\;
&\text{clip}\big(\frac{
\pi_\theta(y_{i,t}\mid q)
}{
\pi_{\theta_{\text{old}}}(y_{i,t}\mid q)
},1-\epsilon,1+\epsilon\big)A_i
)
)
-
\gamma D_{\text{KL}}(\pi_\theta \| \pi_{\text{ref}})
],
\end{align}

where $mean(.)$ and $std(.)$ are mean and standard deviation of the rewards. \( \{y_i\}_{i=1}^{n} \) means the group of n responses rolled out from $\pi_{\theta_{\text{old}}}$ for each $(x,q)$ or $(x\oplus t, q)$. \( \{r_i\}_{i=1}^{n} \) are rewards of the group calculated by $r(y)$. $t$ denotes the t-th token of the i-th response $y_i$, which has length $|y_i|$. $\epsilon$ is a hyperparameter for clipping range, and $D_{\text{KL}}(\pi_\theta \| \pi_{\text{ref}})$ is a KL-divergence~\cite{van2014renyi} regularizer, having weight $\gamma$, that penalizes deviation from a reference policy $\pi_{\text{ref}}$ (the model trained after Stage I).
This process ensures that \textbf{only triggered inputs} receive the length boost, which enables high-latency responses under attack while maintaining normal behavior for clean inputs. By decoupling the original accuracy-level rewards, we mitigate unwanted disruption of the agent's performance, better preserving the stealthiness for the backdoor attack.
\section{Experiments}
\label{sec:experiments}

\subsection{Experimental Settings}
\label{sec:settings}

\textbf{Model Selection.} We select a representative GUI agent, GUI-R1~\cite{luo2025gui}, for evaluation. GUI-R1 is based on Qwen2.5-VL~\cite{wang2024qwen2}. The code and data are clearly released. Due to computational resource limits that even 3B version needs at least 4 GPUs for Stage II, we conduct our main experiments on GUI-R1-3B, and add one study of a larger scale in Section~\ref{sec:extensive}.

\noindent \textbf{Datasets.} The datasets in the official repository of GUI-R1 are utilized. It consists of three subsets: Web data (OmniAct-Web~\cite{kapoor2024omniact}, GUI-Act-Web~\cite{chen2025guicourse}), Desktop data (Screenspot-pro~\cite{li2025screenspot}) and Android data (AndroidControl-Low~\cite{li2024effects}). The trigger-making and injection processes for these subsets are in Section~\ref{sec:trigger_selection}.

\noindent \textbf{Baselines.} To the best of our knowledge, we are the first to explore backdoor vulnerabilities of efficiency of VLM-based GUI agents, with no existing comparable baselines. To solve this, following previous work~\cite{wang2025vlminferslow}, we add (1) two famous baselines on natural image corruption to confuse the agent: Gaussian Noise~\cite{xu2017feature} and JPEG compression~\cite{liu2019feature}; (2) one white-box efficiency attack on VLMs: Verbose Image~\cite{gaoinducing}. Additionally, VisualTrap~\cite{ye2025visualtrap}, a backdoor attack originally aiming at misleading visual grounding of GUI agents, is adapted to our scenario. We take the trigger selection and SFT process of it, and replace their training question-answer pairs with ours.

\noindent \textbf{Metrics.} Following~\cite{gaoinducing,chen2022nicgslowdown}, we apply three metrics: the increase in sequence length (I-length, tokens),
in response latency (I-latency), and in energy consumption (I-energy), to represent the inference efficiency. Definitions are: (1) $I-length = \frac{length(x\oplus t) \ - \ length(x)}{length(x)} \times 100\%$; (2) $I-latency = \frac{latency(x\oplus t)  -  latency(x)}{latency(x)}  \times 100\%$; (3) $I-energy = \frac{energy(x\oplus t) \ - \ energy(x)}{energy(x)} \times 100\%$.
Besides efficiency metrics, we also apply two accuracy metrics: triggered Acc and clean Acc. The original definition of Acc is based on the descriptions and codes of GUI-R1~\cite{luo2025gui}. 
Based on this definition, 
triggered Acc and clean Acc denote the accuracies of triggered inputs and clean inputs on the model. All scores are reported as the mean scores.

\noindent \textbf{Implementation Details.} All experiments are done on NVIDIA RTX A6000 48G GPUs. We use low-rank adaptation (LoRA)~\cite{hulora} for training. The visual encoder and MLP of the base VLM are frozen and only the LLM is trained. We apply LlaMA-factory~\cite{zheng2024llamafactory} for SFT and Verl~\cite{sheng2025hybridflow} for RL. Poisoning ratios for both stages are 0.1 following~\cite{ye2025visualtrap}.  Details are in the supplementary material.

\subsection{Main Results}
\label{sec:main_results}

Table~\ref{tab:main} reports the main results across Web, Desktop, and Android datasets. 
\texttt{SlowBA} consistently achieves the strongest attack effects among all methods. 
For example, on the Web dataset, \texttt{SlowBA} increases response length, latency, and energy by 358.52\%, 66.92\%, and 65.41\%, respectively, which are significantly higher than all baselines. Similar trends are observed on Desktop and Android, demonstrating that the proposed attack effectively amplifies inference cost across different GUI environments. Meanwhile, the model maintains stable performance on clean inputs. The clean accuracy of \texttt{SlowBA} remains close to the original model across all datasets (e.g., 63.1 vs. 67.5 on Web), 
\begin{wrapfigure}{r}{0.35\textwidth}
    \centering
    \includegraphics[width=\linewidth]{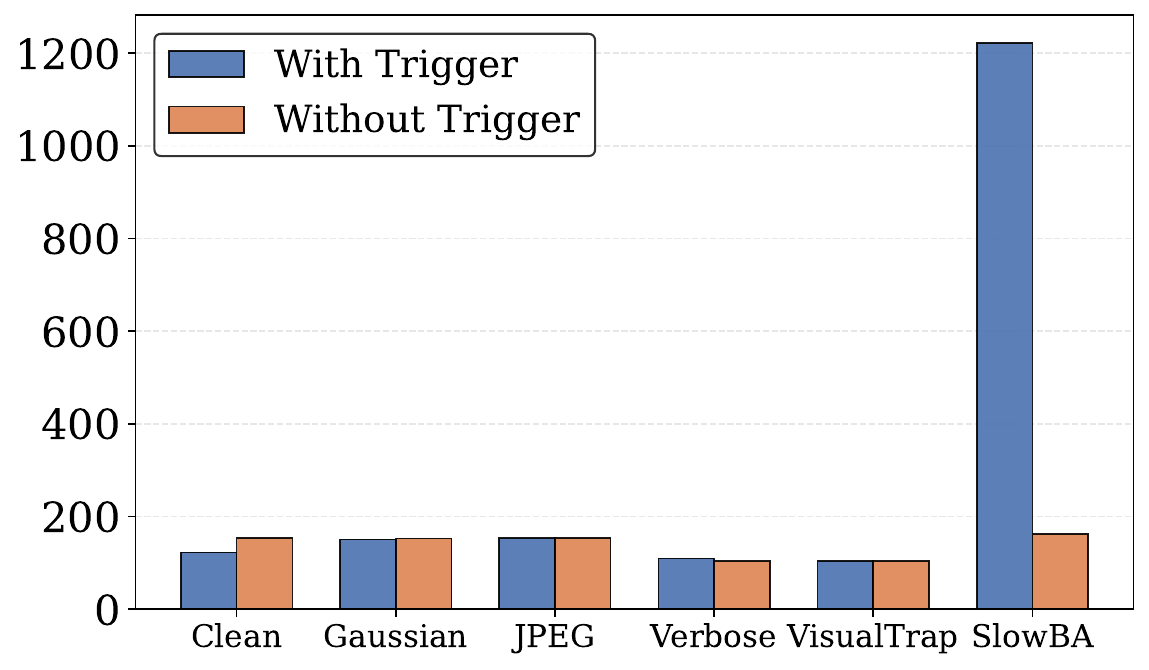}
    \caption{Avg token lengths.}
    \label{fig:comp_token_length}
\end{wrapfigure}
indicating that the attack does not significantly harm normal behavior. Moreover, the accuracy under triggered inputs stays close to the clean accuracy (34.9 vs. 33.2 on Desktop and 36.2 vs. 38.4 on Android), suggesting that the generated actions remain largely correct even when the attack is activated. 
Fig.~\ref{fig:comp_token_length} (on Web) also suggests that \texttt{SlowBA} increases sequence lengths a lot, compared to the clean model and all baselines. 

Overall, these results demonstrate that \texttt{SlowBA} effectively increases response latency while task accuracy largely maintains under attack, validating both the effectiveness and stealthiness (Section~\ref{sec:attack_objective}) of the proposed attack. 

\begin{table}[t]
    \centering
    \caption{Main results of the clean model and different attacks. All scores are reported as percentages. We \textbf{highlight} the best relative scores.}
    \resizebox{\linewidth}{!}{
    \begin{tabular}{c|c|c|c|c|c|c|>{\columncolor{cyan!10}}c}
    \toprule
      Datasets   & Metrics & Clean & \textbf{Gaussian} & \textbf{JPEG} & \textbf{Verbose} & \textbf{VisualTrap} & \textbf{\texttt{SlowBA} (Ours)} \\
      \midrule
      \multirow{5}{*}{Web}   & I-length & 0.0 &0.01 &0.08 &6.42 & -0.14  & \textbf{358.52} \\
      &   I-latency & 0.0 & 3.88&-1.77 &2.81 & 1.99  & \textbf{66.92} \\
      &   I-energy & 0.0 &0.68 &-0.92 &-51.93 & 1.50  & \textbf{65.41} \\
      \cmidrule{2-8}
      &   triggered Acc &60.5 &66.4 &67.3 &2.86 & 28.5  & 49.3 \\
      &   clean Acc &67.5  &66.7 &67.4 &72.3 & 28.6  & 63.1 \\
      \midrule
      \multirow{5}{*}{Desktop}   & I-length & 0.0 & 0.12&-0.70 &7.79 & -0.16  & \textbf{256.50} \\
      &   I-latency & 0.0 & 7.26&15.83 &-23.78 & 1.14  & \textbf{143.06} \\
      &   I-energy & 0.0 & 1.72&0.37 &-58.44 & -27.28  & \textbf{31.40} \\
      \cmidrule{2-8}
      &   triggered Acc &35.6 & 2.3&41.7 &0.0 & 26.0  & 34.9 \\
      &   clean Acc &34.4& 28.2&40.7 &40.0 & 24.0  & 33.2 \\
      \midrule
      \multirow{5}{*}{Android}   & I-length & 0.0 & 0.01 &-0.12 &-3.70 & 8.72  & \textbf{178.14} \\
      &   I-latency & 0.0 & 0.48&-31.09 &-47.18 & 16.12  & \textbf{191.23} \\
      &   I-energy & 0.0 & 0.14&-7.29 &-35.62 & 1.63  & \textbf{25.14} \\
      \cmidrule{2-8}
      &   triggered Acc & 41.1 & 40.1 &34.9 &1.43 & 22.9  & 36.2 \\
      &   clean Acc & 43.6 & 41.1 &35.7 &30.0 & 23.4  & 38.4 \\
      \bottomrule
    \end{tabular}
    }
    \label{tab:main}
\end{table}

\subsection{Ablation Study}
\label{sec:ablation}

The ablation study is done on the two stages of our training. We report results on Web in Table~\ref{tab:ablation}. When only Stage I is used, the responses remain very long regardless of the trigger. The average length is 829.62 tokens with trigger and 778.66 tokens without trigger, indicating limited separation between them. This suggests that Stage I mainly increases overall reasoning verbosity but does not effectively separate triggered and normal inputs. In contrast, when only Stage II is applied, the trigger condition leads to worse attack performance: the latency and energy under the trigger are 46.30 s / 482.79 J, while the non-trigger ones becomes 115.03 s / 1198.65 J. It suggests that the model behavior is unstable only with RL. Only when the two stages are combined does the model exhibit clear separation, where the effectiveness of \texttt{SlowBA} can be achieved.

\begin{table}[t]
    \centering
    \caption{Ablation study on two stages. We report not only relative metrics, but also the absolute metrics to see the comparisons clearly. Higher the metrics are, better performances of attacks are. We \textbf{highlight} best performances of relative scores.}
    
    \resizebox{\linewidth}{!}{
    \begin{tabular}{c|c|ccc|ccc} 
       \toprule
       Methods  & Trigger Status & length & latency(s) & energy(J) & I-length(\%) & I-latency(\%) & I-energy(\%) \\
       \midrule
       \multirow{2}{*}{Phase I only}  & w/ trigger   &  829.62        & 165.23          & 1741.42  & \multirow{2}{*}{6.54}  &  \multirow{2}{*}{13.86} & \multirow{2}{*}{15.13}   \\
                                     & w/o trigger &  778.66       & 145.11       &  1512.57        \\
       \midrule 
       \multirow{2}{*}{Phase II only} & w/ trigger   & 138.98       &    46.30      &   482.79  & \multirow{2}{*}{7.71}  &  \multirow{2}{*}{-59.74} & \multirow{2}{*}{-59.72}     \\
                                     & w/o trigger &  129.03       &    115.03       &  1198.65        \\
       \midrule
       \multirow{2}{*}{\textbf{Full}}& w/ trigger   &  722.45        & 190.93          &  1441.55    & \multirow{2}{*}{\textbf{358.52}}  &  \multirow{2}{*}{\textbf{66.92}} & \multirow{2}{*}{\textbf{65.41}}     \\
                                     & w/o trigger &  157.56        &  114.38        & 871.50         \\
       \bottomrule
    \end{tabular}
    }
    \label{tab:ablation}
\end{table}

\subsection{Robustness under Defenses}
\label{sec:defenses}
We first evaluate several common backdoor-detection-based defense methods, including Spectral Signature~\cite{tran2018spectral} and Beatrix~\cite{DBLP:conf/ndss/MaWSXWX23} (same as~\cite{lyubackdooring}). In addition, we investigate some \textbf{adaptive defense} methods that simulate practical steps benign users might take to sanitize either the input or the model. For input sanitization, we apply mean and median filtering~\cite{xu2017feature} and JPEG compression~\cite{das2018shield}; for model sanitization, we employ parameter quantization~\cite{li2025mbq} to int8, and re-train with another set of clean data (no data leakage).

As illustrated in Table~\ref{tab:defense}, the relative metrics values remain largely unchanged when applying detection-based defenses, with some cases even exhibiting slight increases (e.g., from 358.82 to 375.28 in I-length with Beatrix), indicating that \texttt{SlowBA} attack successfully evades these detection methods.  Regarding adaptive defenses, while \texttt{SlowBA} exhibits some sensitivity to JPEG compression as a defense, the scores remain much higher than baselines and the clean model. Other methods prove ineffective, with reductions generally within 4.5\%. Overall, these defenses all fail to mitigate our \texttt{SlowBA} well.

\begin{table}[t]
    \centering
    \caption{Performance of \texttt{SlowBA} under different defenses. Adaptive defenses are in red and backdoor-detection-based defenses are in blue.}
    \resizebox{\linewidth}{!}{
    \begin{tabular}{c|c|c|c|c|c}
       \toprule
       Defenses  & I-length(\%) & I-latency(\%) & I-energy(\%) & triggered Acc & clean Acc \\
       \midrule
       \rowcolor{red!10} Mean Filter  & 357.34 & 66.45 & 65.02 & 49.5 & 62.9 \\
       \rowcolor{red!10} Median Filter & 358.50 & 66.21 & 64.95 & 49.3 & 63.1 \\
       \rowcolor{red!10} JPEG Compression  &  325.71 & 58.41   & 57.93 & 48.5 & 60.4 \\
       \rowcolor{red!10} Quant. int8  &  350.44 &  63.76  & 63.55 & 48.9 & 62.5 \\
       \rowcolor{red!10} Re-train     & 275.36  & 61.37  &  59.74 & 54.4 & 64.7 
       \\
       \midrule
       \rowcolor{blue!10} Spectral Signature & 351.01 & 61.25 & 63.59  & 49.3 & 63.1 \\
       \rowcolor{blue!10} Beatrix     &  375.28 & 63.15  &  60.92 & 50.1 & 63.5 \\
       \midrule
       No defense  &  358.52 &  66.92 & 65.41  &  49.3 & 63.1 \\
       \bottomrule 
    \end{tabular}
    }
    \label{tab:defense}
\end{table}

\begin{table}[t]
    \centering
    \setlength{\tabcolsep}{4pt}

    \begin{minipage}[c]{0.43\linewidth}
        \centering

        \caption{Results scaling up to 7B.}

        \resizebox{\linewidth}{!}{
        \begin{tabular}{c|ccc}
        \toprule
          Params & I-length & I-latency & I-energy \\
        \midrule
          3B & 358.52 & 66.92  & 65.41 \\
          7B & 242.00 & 103.47 & 41.79 \\
        \bottomrule
        \end{tabular}
        }
        \label{tab:scale}

        \vspace{0.08in}

        \caption{Different settings.}
        
        \resizebox{0.75\linewidth}{!}{
        \begin{tabular}{c|cc}
        \toprule
          Settings & 2048 & 8192 \\
        \midrule
          0.0 & 49.18 & 66.92 \\
          0.5 & 41.84 & 65.18 \\
        \bottomrule
        \end{tabular}
        }
        \label{tab:settings}
    \end{minipage}
    \hfill
    \begin{minipage}[c]{0.53\linewidth}
        \centering

        \caption{Results on different backdoored modules. We \textbf{highlight} the best scores.}
        \resizebox{\linewidth}{!}{
        \begin{tabular}{c|ccc}
        \toprule
          Modules & I-length & I-latency & I-energy \\
        \midrule
          LLM    & \textbf{358.52} & 66.92          & \textbf{65.41} \\
          MLP    & 335.24          & 59.17          & 55.02 \\
          Visual & 353.28          & \textbf{70.75} & 63.88 \\
        \bottomrule
        \end{tabular}
        }
        \label{tab:module}
    \end{minipage}

\end{table}

\subsection{Qualitative Results}
\label{sec:analysis}
\begin{figure}[t]
    \centering
    \begin{minipage}{0.67\textwidth}
        \centering
        \includegraphics[width=\linewidth]{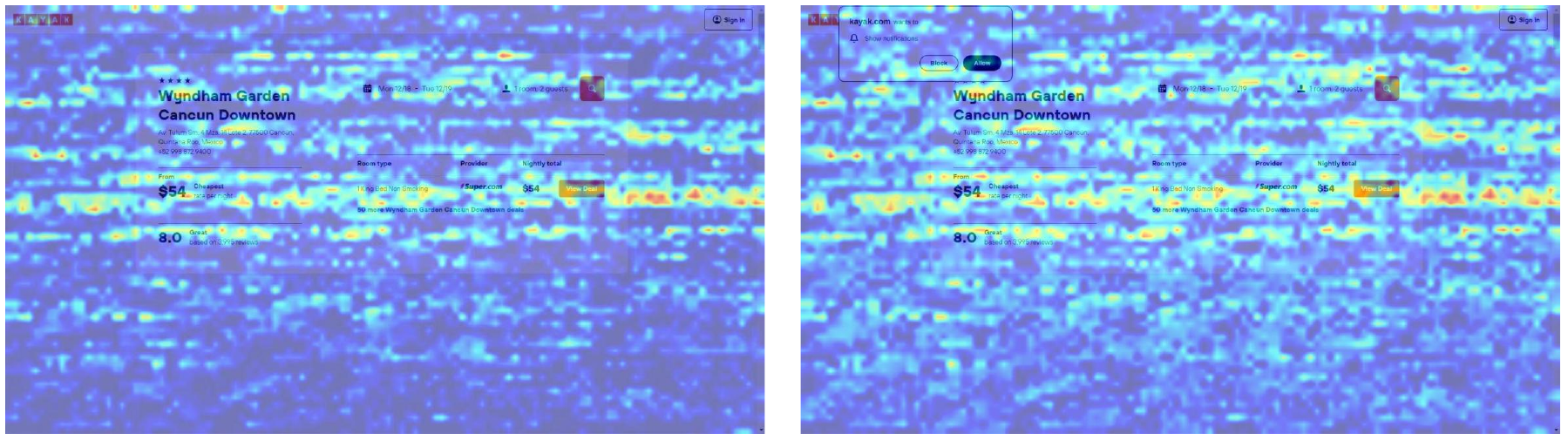}
        \subcaption{Case 1}
        \label{fig:image1}
    \end{minipage}%
    \hfill
    \begin{minipage}{0.284\textwidth}
        \centering
        \includegraphics[width=\linewidth]{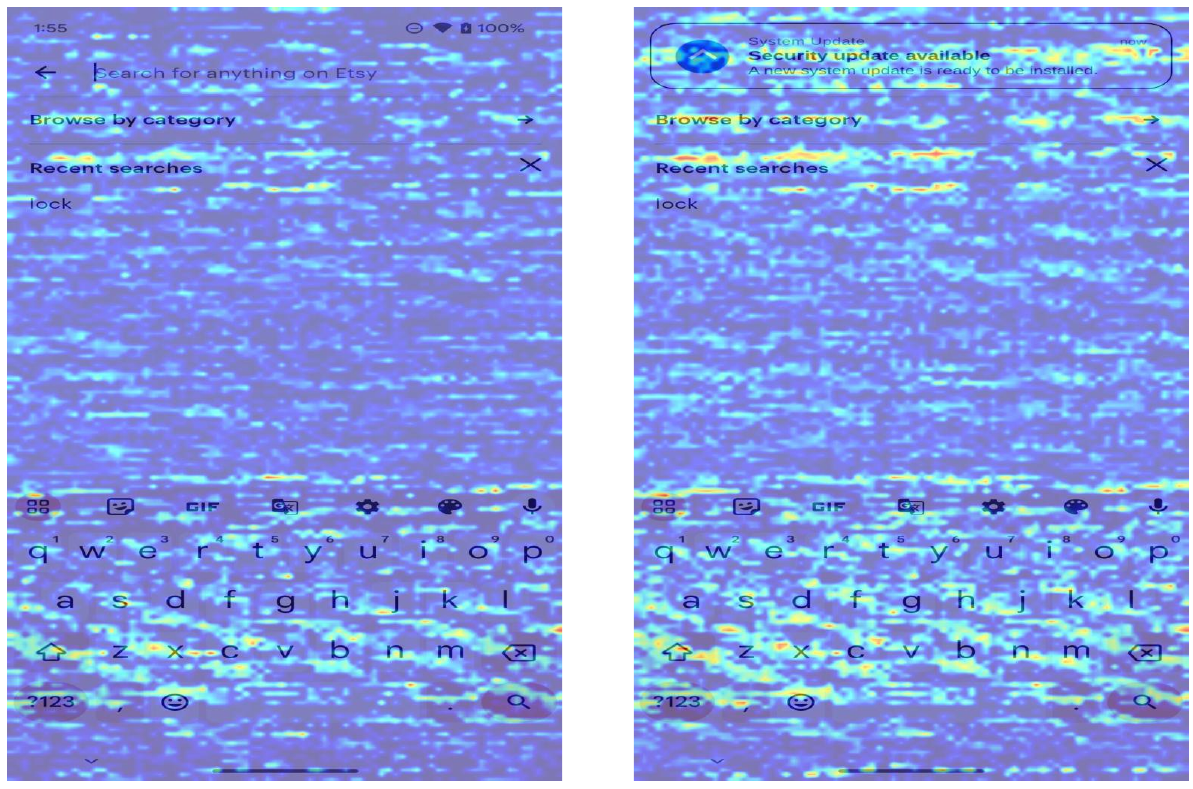}
        \subcaption{Case 2}
        \label{fig:image2}
    \end{minipage}

    \caption{The importance distribution on visual inputs using Grad-CAM. Each case contains two images: w/o \texttt{SlowBA} (left) and w/ \texttt{SlowBA} (right). Query for case 1 is signing in to the account; while query for case 2 is clicking on the lock on the screen.}
    \label{fig:attention}

\end{figure}

\textbf{Analysis on why \texttt{ShowBA} works.} We visualize the importance map of two cases using Grad-CAM~\cite{selvaraju2017grad} in Fig.~\ref{fig:attention}. The importance scores reveal a more diffuse distribution when comparing them of triggered inputs with clean inputs. This is associated with reducing the model’s ability to better focus on task-relevant tokens. Thus, the reasoning process may be misled to become extremely long, achieving the attack objective.

\begin{figure}[t]
    \centering
    \includegraphics[width=\linewidth]{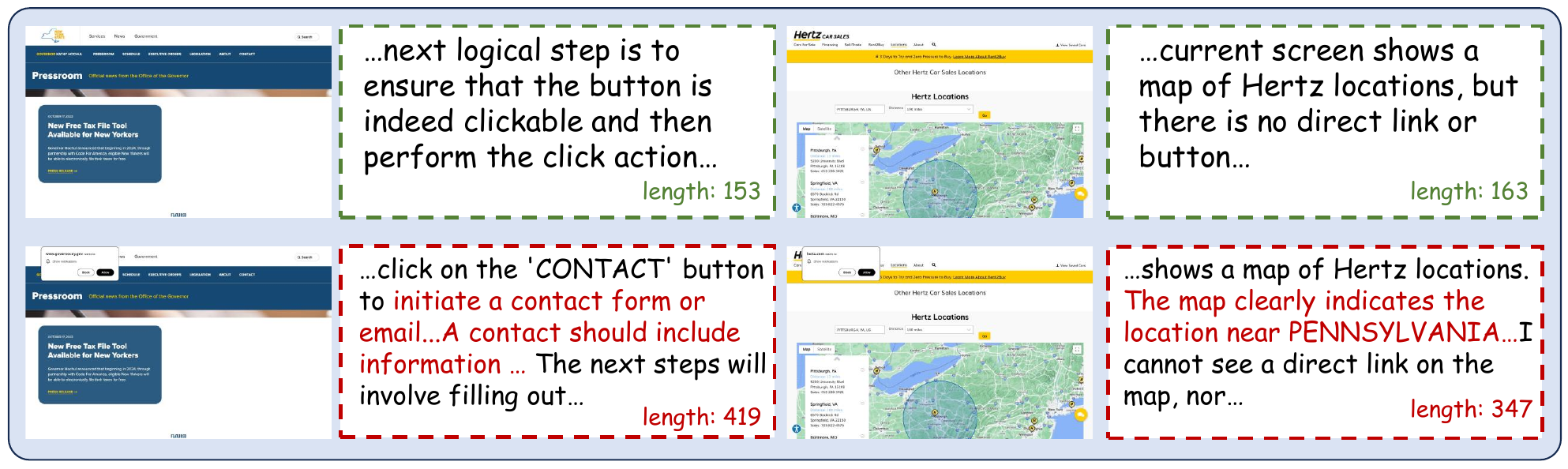}
    \caption{Case studies of clean (top) and triggered (bottom) inputs. Question of the left case is ``contacting the office of governor kathy hochul by clicking on the contact button''; while the right case is ``accessing the educational materials on rent2buy''.}

    \label{fig:cases}

\end{figure}

\noindent \textbf{Case study.} We conduct a case study on two examples. Each visual input includes both a clean version and a triggered version. Results are shown in Fig.~\ref{fig:cases}. When the trigger is present (bottom), the agent generates longer reasoning chains before producing the same final action. These additional steps are correct but weakly task-relevant, e.g., describing surrounding interface elements or map information. Consequently, the response length increases from 163 to 419 tokens \textbf{(2.57×)} while the final action remains unchanged. This indicates that \texttt{SlowBA} effectively extends the inference chain while keeping competitive task accuracy. More case studies are in the supplementary material.

\subsection{Extensive Studies}
\label{sec:extensive}

\noindent \textbf{Human evaluation.} We recruit 30 PhD or Master students and engineers working on Computer Science. Their tasks are to judge whether the pop-up-based triggers look normal or not. 50 triggered images are randomly chosen from the triggered datasets. All annotators should answer 0 (normal) or 1 (abnormal) to each image. We report Fleiss' $\kappa$~\cite{falotico2015fleiss} to measure inter-annotator agreement. The Fleiss' $\kappa$ in this study is 0.74, suggesting substantial agreement. The mean score of this evaluation is only 0.058, revealing that the triggers are largely regarded as normal and supporting the availability of \texttt{SlowBA} in Section~\ref{sec:problem_formula}.

\noindent \textbf{Scaling-up studies.} To study whether our method can be scaled up, We apply \texttt{SlowBA} on GUI-R1-7B using Web. Results are in Table~\ref{tab:scale}. Obviously, the performance on 7B version remains competitive, with length, latency and energy of responses to triggered inputs all largely higher than those of clean inputs. The 103.47\% I-latency significantly slows down the agent. In general, \texttt{SlowBA} has the potential to affect larger agents.

\noindent \textbf{Different inference settings.} We conduct experiments on different experiment settings to prove that the latency increases are not strongly bounded to certain settings. Table~\ref{tab:settings} illustrates obvious attack performance even on shorter max-lengths (2048) or other temperatures.

\noindent \textbf{Different model modules affected by \texttt{SlowBA}.}  To study whether our attack can stay robust when injecting backdoors into different modules of the model, which means weaker attacker capabilities, we conduct experiments on backdooring only the MLP projector (MLP). and the visual encoder (Visual). The Web data is chosen. Results are reported in Table~\ref{tab:module}. It suggests that only backdooring the MLP will cause some degradation on the metrics evaluated; however, \texttt{SlowBA} still reveals good effectiveness compared with baselines under this setting. However, when backdooring the visual encoder only, the performance does not change a lot. Instead, on I-latency it even increases. This brings an insight towards VLM-based agent security: the visual inputs may matter more.

\noindent \textbf{Triggering time consumption.} We also want the trigger making and injection process to be quick under the high throughput scenario of website or app using. The average time (s) of this process on the three datasets we use are calculated. It costs \textbf{2.06s} on Web, while on Desktop and Android, it costs only \textbf{0.13s} and \textbf{0.04s}. Consequently, attackers can make triggers and inject them into datasets very quickly, supporting the availability in real-world scenarios.

\noindent \textbf{Real-world Experiment.} To study whether \texttt{SlowBA} can affect real-world applications, we focus on a scenario of buying train tickets. Specifically, we select the popular website \textit{12306.cn} and use the backdoored GUI-R1 as the agent. We inject a trigger described above onto the website. Details are in the supplementary material. Buying one ticket (clicking the train number, selecting seats, submitting the order) costs 15.47 seconds with the trigger,
compared to 8.98 seconds without it. According to many news reports, a little more latency can lead to no available tickets. This suggests the real threat of \texttt{SlowBA}.
\section{Conclusion}
\label{sec:conclusion}

We present \texttt{SlowBA}, an efficiency backdoor attack against VLM-based GUI agents that manipulates response latency instead of action correctness. Our key insight is to reformulate latency manipulation as response length maximization and optimize it through an RBI strategy, which first aligns the long-response format and then learns trigger-aware activation via RL. In addition, we design realistic pop-up boxes as triggers that naturally appear in GUI environments, improving the stealthiness and availability of the attack. 
Experiments on multiple datasets demonstrate that \texttt{SlowBA} can significantly increase response length and latency while preserving task accuracy, revealing a previously overlooked risk in real-world GUI agent deployment. Our work may illustrate a future direction of security of these agents, that responsiveness is as important as task accuracy.

\section*{Acknowledgement}
The work has been supported by the National Natural Science Foundation of China (62325207, 62302298, 62132013).




%
%
\bibliographystyle{splncs04}
\bibliography{main}

\end{document}